\documentstyle[prd,aps,floats,psfig]{revtex}
\def\lsim{\mathrel{\lower2.5pt\vbox{\lineskip=0pt\baselineskip=0pt
\hbox{$<$}\hbox{$\sim$}}}}
\def\gsim{\mathrel{\lower2.5pt\vbox{\lineskip=0pt\baselineskip=0pt
\hbox{$>$}\hbox{$\sim$}}}}

\newcommand{\ima}{{\mbox{Im}\,}}

\newcommand{\be}{\begin{equation}}
\newcommand{\ee}{\end{equation}}

\newcommand{\NP}[1]{Nucl.\ Phys.\ {#1}}

\newcommand{\PL}[1]{Phys.\ Lett.\ {#1}}

\newcommand{\PR}[1]{Phys.\ Rev.\ {#1}}
\newcommand{\PRL}[1]{Phys.\ Rev.\ Lett.\ {#1}}

\begin{document}
\draft
\renewcommand{\topfraction}{0.8}
\twocolumn[\hsize\textwidth\columnwidth\hsize\csname
@twocolumnfalse\endcsname

\preprint{hep-ph/01XXXXXX}
\title{Chiral Perturbation Theory and the $f_2(1270)$ resonance.}
\author{A. Dobado and
 J. R. Pel\'aez}
\address{Departamento de F{\'\i}sica Te{\'o}rica I and II,
  Universidad Complutense de Madrid, 28040-- Madrid,\ \ Spain}
\date{November, 2001}
\maketitle

\begin{abstract}
Within Chiral Perturbation Theory,
we study elastic pion scattering in the $I=0, J=2$, channel,
whose main features are the $f(1270)$ resonance
and the vanishing of the lowest order.
By means of a chiral model that includes an explicit resonance
coupled to pions, we describe the data and
calculate the  resonance contribution to the $O(p^4)$ and
$O(p^6)$ chiral parameters. We also generalize
the Inverse Amplitude Method to higher orders ,
which allows us to study channels with vanishing lowest order.
In particular, we apply it to the  $I=0,J=2$ case, finding a good
description of the $f_2(1270)$ resonance, as a pole
in the second Riemann sheet.
\end{abstract}


\vskip2pc]

Chiral Perturbation Theory (ChPT) \cite{chpt1,books} is a powerful
tool to describe low energy hadronic interactions. ChPT 
is based on the identification
of pions, kaons
and the eta as the Goldstone Bosons associated to the spontaneous
chiral symmetry breaking of QCD (pseudo-Goldstone bosons indeed, since the 
three lightest have a small mass). The ChPT Lagrangian is then 
built as a derivative and mass expansion over the symmetry breaking scale 
$4\pi F\simeq 1.2 \hbox{GeV}$, compatible with the 
symmetry constraints. The calculations are renormalizable 
order by order and depend on just a finite set of parameters
at each order, which  can be determined from a few experiments 
and then used to obtain predictions for other processes.
These parameters contain information on heavier states not
included explicitly in the Lagrangian \cite{chpt1,saturation,donoghue}.

In this work, and  within the context of $SU(2)$ ChPT (the $u$ and $d$
quark sector), we first study the contribution  from the $I=0,J=2$ 
lightest resonance
to the chiral parameters, using a resonant model that describes well 
the data on that channel. Next, we study how, by means
of unitarization methods, it is possible to 
generate a resonance from the ChPT expansion in the $I=0,J=2$ channel.
Although these techniques, and particularly the Inverse Amplitude Method, have
been extensively applied in the literature, obtaining remarkable descriptions
of meson-meson scattering, they had never been applied to this channel. The reasons
were that the lowest chiral order vanishes, so that the formalism has to be 
generalized, and that the first contribution to the imaginary part of the
amplitude appears at three loops, where there are no calculations available.
We conclude by showing and discussing our numerical results
confronted with data.

\vspace{-.5cm}
\subsection*{$f_2(1270)$ contribution to the chiral parameters}
\vspace{-.3cm}
\label{resonance}
 The most remarkable feature of the D wave isoscalar pion
 scattering channel is the $f_2(1270)$. Therefore a possible
 phenomenological approach to this channel is a model
where a $J=2, I=0$ resonance is introduced explicitly in a chiral invariant
way. In a first approximation we will neglect the kaons since their
branching ratio from the $f_2(1270)$ is about 5\%.
Thus we consider the $SU(2)$ chiral symmetry framework,
where the pions are grouped in $U(x)=\exp (i \tau^a\pi^a(x)/F)$,
$\tau^a$ being the Pauli matrices. The $f_2$
is described by a symmetric real tensor field
$f_{\mu\nu}$ with perturbative mass $M$. Its chiral
invariant interaction to pions at lowest order in derivatives
is \cite{donoghue}
\begin{eqnarray*}
{\cal L}_{int}=g f_{\mu\nu}tr \partial^\mu U \partial^\nu
U^\dagger,
\end{eqnarray*}
where $g$ is a coupling with dimension of energy. Note that
the resonance only couples to an even number of pions,
and in particular the interaction with two pions is
\begin{eqnarray*}
{\cal L}_{int}=\frac{2 g}{F^2}f_{\mu\nu} \partial^\mu \pi^a
\partial^\nu   \pi^a.
\end{eqnarray*}
In order to obtain the Feynman rules, let us recall that
the resonance in the initial or final states is described by a
plane wave $\Phi_{\mu\nu}(\vec k,\lambda)$, where $\vec k$ is the
momentum and $\lambda$ is the helicity, satisfying:
$(k^2-M^2)\Phi_{\mu\nu}(\vec k,\lambda)=0$.
The sum over polarizations is given by \cite{Novozhilov}
\begin{eqnarray*}
\sum_{\lambda}\Phi_{\mu\nu}(\vec k,\lambda)\Phi_{\rho\sigma}(\vec
k,\lambda)^\dagger =X_{\mu\nu,\rho\sigma},
\end{eqnarray*}
with $X_{\mu\nu,\rho\sigma}=\frac{1}{2}(X_{\mu\rho}X_{\nu\sigma}+
X_{\mu\sigma}X_{\nu\rho})-\frac{1}{3}(X_{\mu\nu}X_{\rho\sigma})$
and
\begin{eqnarray*}
X_{\mu\nu}=g_{\mu\nu}-\frac{k_{\mu}k_{\nu}}{M^2},
\end{eqnarray*}
Thus, the $f_2\rightarrow\pi\pi$ decay amplitude is
\begin{eqnarray*}
T(f \rightarrow \pi\pi)=-\frac{4g}{F^2}k_1^\mu k_2^\nu
\Phi_{\mu\nu},
\end{eqnarray*}
and its partial  width is
\begin{eqnarray*}
\Gamma(f \rightarrow \pi\pi)=\frac{g^2}{80\pi M^2
F^4}(M^2-4M_\pi^2)^{5/2}.
\end{eqnarray*}
Using \cite{PDG}
 $M_{f_2}=1270 \hbox{MeV}\simeq M$,
$\Gamma(f \rightarrow \pi\pi)= 158 \hbox{MeV}$,
$F=92.4 \hbox{MeV}$ and $M_\pi=139.57 \hbox{MeV}$, we find
 $g\simeq 43.3$.
It is also straightforward to obtain
the pion scattering amplitude from this model, which,
due to crossing and chiral symmetry, has the following form
\begin{eqnarray*}
T_{abcd}(s,t,u)&=&A(s,t,u)\delta_{ab}\delta_{cd}  \nonumber   \\
&&+A(t,s,u)\delta_{ac}\delta_{bd}+A(u,t,s)\delta_{ad}\delta_{cb}.
\end{eqnarray*}
where $a,b,c$ and $d$ are the isospins of the pions
and $s$, $t$ and $u$ are the Mandelstam variables. In our
model the three terms correspond to the $f_2$ exchange in the
$s$, $t$ and $u$ channels, respectively, with its
propagator given by
\begin{eqnarray*}
D_{\mu\nu,\rho\sigma}(k)=i\frac{X_{\mu\nu,\rho\sigma}}{k^2-M^2},
\end{eqnarray*}
so that
\begin{eqnarray}
\label{Afper} A_f(s,t,u)=\frac{g^2}{F^4(M^2-s)}\left\{ 2[(2 M_\pi^2-t
)^2+(2M_\pi^2-u)^2] \right.\nonumber \\
\left.\hspace{-.1cm}-\frac{4}{3}\left(s-2M^2_\pi\right)^2-\frac{2\,s^2}{3\,M^2}
(s+4M_\pi^2)+\frac{2}{3}\left(\frac{s^2}{M^2}\right)^2\right\}.\hspace{.5cm}
\end{eqnarray}
In order to compare with ChPT we calculate the lowest order in the momenta
and the pion mass, and we find
\begin{eqnarray*}
 A_f(s,t,u)&=&\frac{g^2}{F^4M^2}\left\{ 2[(2 M_\pi^2-t
)^2+(2M_\pi^2-u)^2]\right.\nonumber    \\
&&\left. -\frac{4}{3}\left(s-2M^2_\pi\right)^2
\right\}+O(p^6).
\end{eqnarray*}
As expected the $O(p^2)$ vanishes. By
comparing with the ChPT scattering
amplitude \cite{chpt1}, we obtain
the $f_2$ contribution to the chiral
parameters $\bar l_1$ and $\bar l_2$
\begin{eqnarray*}
\Delta \bar l_2 =  -\frac{3}{2} \Delta \bar l_1 =
\frac{96 \pi^2 g^2
}{M_f^2},
\end{eqnarray*}
in agreement with the calculation in \cite{donoghue}
performed with a different notation and in the chiral limit.
Thus $ \Delta \bar l_1 \simeq-0.65$ and
$ \Delta \bar l_2 \simeq 0.95$. There is no
contribution to $\bar l_3$ and $\bar l_4$. Let us now
compare with the dominant $\rho(770)$ contribution\cite{chpt1}
\begin{eqnarray*}
\Delta \bar l_1 = -2\Delta \bar l_2= -\frac{96 \pi^2 f^2 }{M_\rho^2},
\end{eqnarray*}
so that $\Delta \bar l_1 = -7.6$ and $\Delta \bar l_2 = 3.8$, since
the $\rho\pi\pi$
 chiral invariant coupling is $f\simeq 69 \hbox{MeV}$ \cite{chpt1}.
Nevertheless, the $f_2$ contributions are comparable to
those of scalar resonances \cite{saturation,donoghue}.
At $O(p^6)$, $\pi\pi$ scattering can be parametrized with 
six constants $b_i$ \cite{bijnens}. The first four are
dominated the $\bar l_i$, and only $b_5$ and $b_6$
are genuinely $O(p^6)$, whose $f_2$ contribution is:
\begin{eqnarray*}
\Delta  b_5 = -\Delta b_6 = -\frac{F^2 g^2 }{M_f^4}\simeq -6.8\times 10
^{-6},
\end{eqnarray*}
whereas that of the $\rho$ is
\begin{eqnarray*}
\Delta  b_5 & = &    \frac{1}{3}\Delta b_6
=\frac{F^2 f^2 }{4 M_\rho^4}\simeq 3\times 10
^{-5}.
\end{eqnarray*}
When obtaining the  $f_2$ contribution to  the chiral parameters
from our the model, we may wonder how well it describes the $J=2, I=0$
data. To that end we have to evaluate the partial wave
\begin{eqnarray*}
a_{02}(s)=\frac{1}{64\pi}\int_{-1}^1\,d(\cos\theta)
T_{I=0}(s,t,u)P_{J=2}(\cos\theta),
\end{eqnarray*}
where $T_0(s,t,u)=3A(s,t,u)+A(t,s,u)+A(u,t,s)$.
However the amplitude in eq.(\ref{Afper}) is not appropriate
since it is a perturbative amplitude where the resonance
appears with zero width (indeed, it is singular at $s=M^2$).
Frequently, this problem is solved introducing
by hand the width in the resonant propagator. Such an amplitude
behaves as a Breit-Wigner  around
the resonance position, but this method does not provide the proper
analytic structure. In addition, it usually
breaks chiral symmetry and in particular spoils
 the Weinberg low energy theorems.
Therefore we will consider here a different method with better properties.
Instead of the tree level partial wave $a_{02}$ obtained from
Eq.\ref{Afper} we will use
\begin{equation}
\tilde  a_{02}(s) =\frac{a_{02}(s)}{1-J(s)a_{02}(s)},
\label{aconj}
\end{equation}
where $J(s)$ is the Mandelstam two body function
\begin{eqnarray*}
J(s)=\frac{\sigma(s)}{\pi}\log \frac{\sigma(s) -1}{\sigma(s) +1}
\quad
\hbox{where}
\quad
\sigma (s)=\sqrt{1-\frac{4 M_\pi^2}{s}}.
\end{eqnarray*}
Note that eq.(\ref{aconj}) is nothing but
the resummation of a
geometric series generated by the tree level amplitude and the two
pion s-channel loop. Thus it is not perturbative in the $g$
coupling. Its main properties are the following:
it has a right cut  starting at $s=4 M_\pi^2$
and there is a pole in the second Riemann sheet associated
to the resonance,
which provides the usual Breit-Wigner shape.
Similar approaches with other resonances have been
successfully applied in \cite{eulogio}.
As usual, the physical mass and the total width of the resonance can
be obtained approximately from the position of the pole in the complex plane.
Finally $\tilde a_{02}$ is unitary, i.e., for $s>4 M_\pi^2$, it
satisfies
\begin{equation}
\ima \,\tilde a_{IJ}= \sigma \mid \tilde a_{IJ}  \mid  ^2\quad\Rightarrow\quad
\ima \,\tilde a_{IJ}^{-1}=-\sigma
\label{uni}
\end{equation}
In Fig.1, we show a fit to the data of the $\tilde a_{02}$ phase,
with parameters $M=1157\hbox{MeV}$ and $g=44\hbox{MeV}$,
 in good agreement with the estimations
presented above. Up to here we have obtained information
on the ChPT parameters from an explicit resonance. Let us now
see how we can also generate a resonance from ChPT.

\vspace{-.3cm}
\subsection*{The Inverse Amplitude Method for channels with
vanishing lowest order}
\vspace{-.3cm}
Within ChPT  the amplitude is an expansion
\begin{equation}
a_{ChPT}(s)=a_2(s)+a_4(s)+a_6(s)+a_8(s)+...
\label{chiralexp}
\end{equation}
where $a_k(s)$ stands for the $O(p^k)$ contribution,
or more precisely, the $O(1/F^k)$ term. For  $s>4 M_\pi^2$,
$a_{ChPT}$ only satisfies unitarity, eq(\ref{uni}),
in a perturbative sense:
\begin{eqnarray}
 \ima a_2 & = & 0,    \qquad \quad \ima a_6 = a_2\sigma a_4^* +a_4 \sigma a_2,
 \label{pertuni} \\
 \ima a_4 & = & a_2 \sigma a_2, \quad  \ima a_8  =  a_2
\sigma a_6^* + a_6 \sigma a_2 + a_4 \sigma a_4^*,\quad
 ...\nonumber
\end{eqnarray}
In the elastic region, all the contributions to
$\ima a_{ChPT}$ come from the two pion loop function $J(s)$.
Indeed, by analyzing the different Feynman diagrams
contributing to eq.(\ref{chiralexp}) it is possible to write
$a_4=a_2\,J\,a_2+a_{4L}$,
where $a_{4L}$ contains the polynomial and the left cut
contribution to $a_4$ but is real when $s>4M_\pi^2$.
This is due to the fact that we can only get the imaginary part from
one loop (one $J(s)$) with two vertices of $O(p^2)$ (the two $a_2$ factors),
and the rest has to be real if $s>4M_\pi^2$.
Similarly
\begin{eqnarray}
a_6&=&a_2Ja_2Ja_2+2a_{4L}Ja_2+a_{6L}, \nonumber\\
a_8&=&a_2Ja_2Ja_2Ja_2+3a_{4L}Ja_2Ja_2  \nonumber \\
&&+2a_{6L}Ja_2+a_{4L}Ja_{4L}+a_{8L}, \label{a6a8}
\end{eqnarray}
where the $a_{kL}$ include the corresponding left cut and polynomial
contributions, are  renormalization scale  independent and real
on the right cut. Note that $\ima J(s)=\sigma (s)$ for $s>4M_\pi^2$
so that eqs.(\ref{pertuni}) follow immediately.

A case of special interest for this work occurs when
$a_2=0$. Then  $a_{4L}=a_4$  and $\ima a_4=0$ in
the physical region. The same happens with $a_6$. Thus the first right cut
contribution comes from  $a_8$, which satisfies
\begin{equation}
a_8=a_4Ja_4+a_{8L}, \quad
\ima a_8= a_4 \sigma a_4, \quad\forall s>4M_\pi^2,
\label{im8}
\end{equation}
Actually, this occurs for $I=0, J=2$  where
the chiral expansion starts at $O(p^4)$, i.e., $
a=a_4+a_6+a_8+...$

In order to improve the unitary behavior of the chiral expansion
one of the most widely used techniques is the
Inverse Amplitude Method (IAM). Its name is due to the fact that
$a_{IJ}^{-1}$ has the same analytic
structure as $a_{IJ}$ (apart from eventual
new poles coming from the amplitude zeros). In particular,
it should also have a right cut on $s>4M_\pi^2$,
where $\ima a^{-1}=-\sigma$ due to  eq.(\ref{uni}).

The IAM can be derived with the help of an auxiliary
function $G\equiv a_2^2 a^{-1}$ which
satisfies a dispersion relation with exactly the same right cut contribution
as $a_2-a_4$, since $a_2$ is real
and $\ima a^{-1}=-\sigma$.
Indeed, if the left cut contribution and the polynomial part of $G$
are evaluated perturbatively we arrive to $G\simeq a_2-a_4$. Hence
we find a unitarized
amplitude  $\tilde a=a_2^2/(a_2-a_4)$. The details of this
derivation can be found in \cite{IAM1}. Let us simply recall
that this simple formula has been applied to the
$\pi\pi$ and $\pi K$ elastic scattering \cite{IAM1}, and it  generates
 the $\sigma$, $\rho$ and  $K^*$ resonances from the
corresponding $O(p^4)$ amplitudes. A similar equation in matrix form,
 although without a justification from
dispersion theory, has also been applied within a coupled
channel formalism, describing successfully all the meson-meson interactions
below 1200 MeV and generating seven light resonances \cite{IAM2}.
In addition, this method has also been generalized
both to $O(p^6)$ calculations for the lowest spin channels where $a_2\neq0$
\cite{IAM1,Juan}.

However, the IAM has not been derived or applied when $a_2=0$. In what follows,
we will present a generalization of the IAM equation and its derivation
for the case when $a_2=0$. In particular we will apply the method to the 
$I=0, J=2$
channel. In this case it is possible to write a dispersion relation for 
the chiral expansion up to $O(p^8)$ (with five subtractions to ensure 
convergence). As
discussed before, the first non-vanishing contribution to $\ima a$
on the right cut comes from $a_8$, eq.(\ref{im8}). Thus the right cut
contribution to this dispersion relation is
\begin{eqnarray*}
a_{8R}(s)=\frac{(s-s_0)^5}{\pi}\int_{4M_\pi^2}^{\infty}
\frac{a_4(s')\sigma(s')a_4(s') ds'}{(s'-s_0)^5(s'-s-i\epsilon)}
\end{eqnarray*}
where we have used the second relation in eq.(\ref{im8}) and $s_0$
is a subtraction point. This strongly suggest the use of the
auxiliary function $G\equiv a_4^2 a^{-1}$,
since $\ima G= - \ima a_8$ on the right cut. Writing another dispersion
relation (with five subtractions) for $G$, its right cut contribution
will be precisely $-a_{8R}$. Neglecting the possible pole
contribution and evaluating the left cut and the
polynomial contributions perturbatively it is not hard to find
\begin{equation}
G \simeq a_4-a_6-a_8+\frac{a_6^2}{a_4},\,\Rightarrow\,\tilde a
=\frac{a_4}{1-\frac{a_6}{a_4}-\frac{a_8}{a_4}+\frac{a_6^2}{a_4^2}}
\label{IAMfora20}
\end{equation}
This is the generalized expression for the IAM,
which is exactly unitary and has the correct low energy expansion
required by ChPT, i.e., eq.(\ref{chiralexp}) with $a_2=0$.

Alternatively, the unitarized
amplitude above could be derived by considering the $[2,2]$
Pad\'e approximant of the chiral expansion in $1/F^2$, namely
\begin{eqnarray*}
a^{[2,2]}=\frac{a_2a_4-a_2^2a_6+a_4^3-2a_2a_4a_6+a_2^2a_8}
{a_4^2-a^2a^6+a_2a_8-a_4a_6+a_6^2-a_4a_8
}
\end{eqnarray*}
 and then setting $a_2=0$
\vspace{-.4cm}
\subsection*{Results and conclusion}
\vspace{-.3cm}
In what follows we are going to confront  eq.(\ref{IAMfora20})
with the $ I=0, J=2$ scattering data. However, at present there is no
calculation of $a_8$ available, and probably it will remain unavailable for
a long time. However from  eq.(\ref{a6a8}) we see that only
$a_{8L}$ is unknown. Since we will be interested on the resonant
region, $\sqrt{s}\simeq 1200 \hbox{MeV}$, we can expect 
that the left cut logarithmic
contribution will be small. Concerning the polynomial, we also expect
the dominant term to be $a_{8L} \sim c\,s^4$,
 since any other polynomial term will be suppressed
by powers of $M_\pi^2/s$. Since ChPT is
an expansion in powers of momenta over $4\pi F$ we get a 
crude estimate of $c\simeq (1/4\pi F)^8\simeq 3\times 10^{-25}\,
\hbox{MeV}^{-8}$.

Thus, in Fig.1 (dashed line) we compare the $ I=0, J=2$
phase shift data with the results of applying eq.(\ref{IAMfora20}) to the
ChPT amplitude with the parameters listed as set I in Table I.
It is possible to get a remarkable description of the experiment, but it is not
so good when comparing with the values given in the literature, 
listed in column two
and three of the same table. Nevertheless, our parameters have the correct 
order of magnitude.
Let us also remark that there are just six free parameters up to $O(p^6)$, 
with rather large uncertainties.
\begin{figure}
\hspace*{-.5cm}
\hbox{\psfig{file=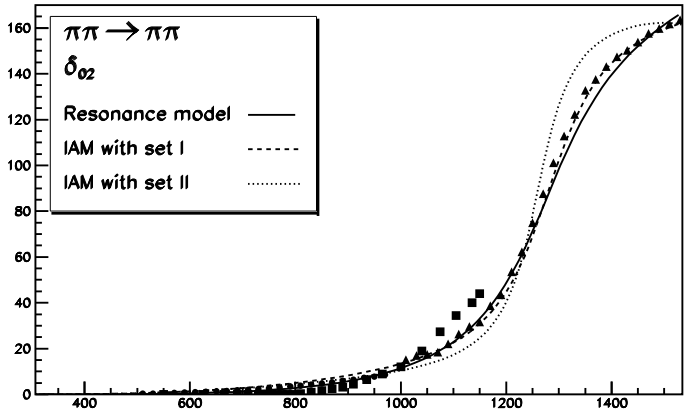,width=9cm}}
\vspace{-.2cm}
\caption{Experimental data ( [12] squares,[13] circles,
[14] triangles)
on the $,I=0, J=2$ $\pi\pi$ scattering
phase shift and the results of
our unitarized chiral resonance model (continuous line), 
of eq.(8)
with the parameters in set I (dashed line), and with those 
in set II (dotted line)}
\end{figure}
However, such an accurate description of the data is 
somewhat unrealistic. The reason
is that in our approach we are only taking into account 
the $\pi\pi$ state, whereas
the actual $f_2$ resonance has 8\% and 7\% branching 
ratios to four pions and 
two kaons,
respectively. Just with pions, we should expect to get a 15\% 
narrower resonance. 
We show
in Fig.1, as a dotted line, the result of applying 
eq.(\ref{IAMfora20}) with set II in Table I. These parameters 
are compatible with
those given in the literature, and from Fig.1, we see that, actually, 
they yield a narrower
resonance, consistently with our expectations. The introduction 
of the $K\bar{K}$
or the $4\pi$ coupled states is beyond the scope of this work.

The final consistency check is the value of the $c$ parameter. As a 
matter of fact we have
found many sets of parameters yielding results either like the dashed or 
the dotted lines
in Fig.1. For all of them, $c\simeq 10^{-25}$ to $10^{-24}\,\hbox{MeV}^{-8}$,
in good agreement with our expectations. The consistency of the whole 
picture is more remarkable taking into account the 
experimental errors, not given in the original references, but that could  
be roughly estimated by comparing the difference on the data points 
in the overlapping region of different experiments (see Fig.1
between 1000 and 1200 MeV).

\begin{table}
    \begin{tabular}{|c|c|c||r|r|}
&$O(p^6)$ ChPT [8]& $O(p^6)
$ IAM [11] & set I & set II\\ \hline
$10^2b_1$ & -9.2 ... -8.6 &$-7.7\pm1.3$ & -3 & -6.6 \\ \hline
$10^2b_2$ &  8.0 ... 8.9 &$7.3\pm0.7$ & 4 & 6.4 \\ \hline
$10^3b_3$ &  -4.3 ... -2.6 &$-1.8\pm1.6$ & 3.8 & -3.6 \\ \hline
$10^3b_4$ &  4.8 ... 7.1 &$4.8\pm0.1$ & 7 & 6.7 \\ \hline
$10^4b_5$ &  -0.4 ... 2.3 &$1.3\pm0.2$ & 8.7 & 4.0 \\ \hline
$10^4b_6$ &  0.7 ... 1.5 &$0.2\pm0.2$ & 1.6 & 1.5 \\
    \end{tabular}
    \caption{ Estimates of the $O(p^6)$ parameters
are given in column two. In the third column we
 give values that,
with the IAM up to $O(p^6)$, fit very well 
$\pi\pi$ scattering in
the $(I,J)= (0,0), (1,1)$ and $(2,0)$ channels.
Set I with eq.(8)
describes remarkably well the $I=0,J=2$ data, but only agrees 
in the order of magnitude with previous values. Set II 
agrees better
with [11], but yields a narrower resonance
(see Fig.1), due to other coupled states not present
in our approach. }
\end{table}

In summary, in this work 
we have studied 
a chiral model with a $I=0, J=2$ resonance which 
describes the $\pi\pi$ scattering data on that channel.
We have then calculated the resonance contributions
to the chiral parameters that govern $\pi\pi$ scattering
at one and two loops, finding that, as expected, they are 
subdominant with respect to those of vector mesons 
(that is vector meson dominance), but comparable with
the contributions from scalar resonances. We have also given a generalization
of the Inverse Amplitude Method to higher orders,
which, in particular, is applicable to channels with vanishing lowest order.
When applied to the $I=0,J=2$ channel, the IAM is able to generate
a resonant behavior from the chiral expansion,
in agreement with the data, taking
into account that we are only considering the two pion state.
This is an illustration of the power of this unitarization method
which still gives qualitative
 results even close to its applicability limits.

\vspace{-.5cm}
\subsection*{Acknowledgments.}
\vspace{-.4cm}

Work supported from the Spanish CICYT projects
FPA2000-0956, PB98-0782 and BFM2000-1326.

\end{document}